\documentclass[conference]{IEEEtran}
\IEEEoverridecommandlockouts

\usepackage{cite}
\usepackage{amsmath,amssymb,amsfonts}
\usepackage{algorithmic}
\usepackage{graphicx}
\usepackage{textcomp}
\usepackage{xcolor}
\usepackage{caption}    
\usepackage{subcaption} 
\usepackage{enumitem}
\usepackage{booktabs}
\def\BibTeX{{\rm B\kern-.05em{\sc i\kern-.025em b}\kern-.08em
    T\kern-.1667em\lower.7ex\hbox{E}\kern-.125emX}}
\usepackage{scalerel}

\newcounter{takeaway}

\newcommand{\pimodel}{$\pi_{0.5}$ }

\begin{document}
\bstctlcite{IEEEexample:BSTcontrol}


\title{Analysis of Multi-Client Inference Techniques for Vision-Language-Action Models}






\author{Anonymous Author(s)}

\author{\IEEEauthorblockN{Stefan de Lasa}
\IEEEauthorblockA{\textit{University of Toronto}}
\and
\IEEEauthorblockN{Mohammadreza Mofayezi}
\IEEEauthorblockA{\textit{University of Toronto}}
\and
\IEEEauthorblockN{Gennady Pekhimenko}
\IEEEauthorblockA{\textit{University of Toronto}}
\and
\IEEEauthorblockN{Nandita Vijaykumar}
\IEEEauthorblockA{\textit{University of Toronto}}
}

\maketitle

\begin{abstract}
Vision-Language-Action models (VLAs) are a new class of machine learning model for generalizable robot control. While they enable robots to perform complex tasks in unseen environments, serving VLAs leaves hardware idle between inference passes and under-utilized during inference. Serving multiple clients at once can improve utilization, but fundamental multi-client techniques such as batching and co-execution have not been characterized for VLAs. We therefore characterize the trade-offs these known techniques impose across latency, throughput, power, and hardware utilization. Our findings inform trade-offs and challenges for future multi-VLA serving systems.
\end{abstract}

\section{Introduction}
\begin{figure}[h]
    \centering
    \begin{minipage}{0.49\columnwidth}
        \centering
        \vstretch{.96}{\includegraphics[width=\columnwidth]{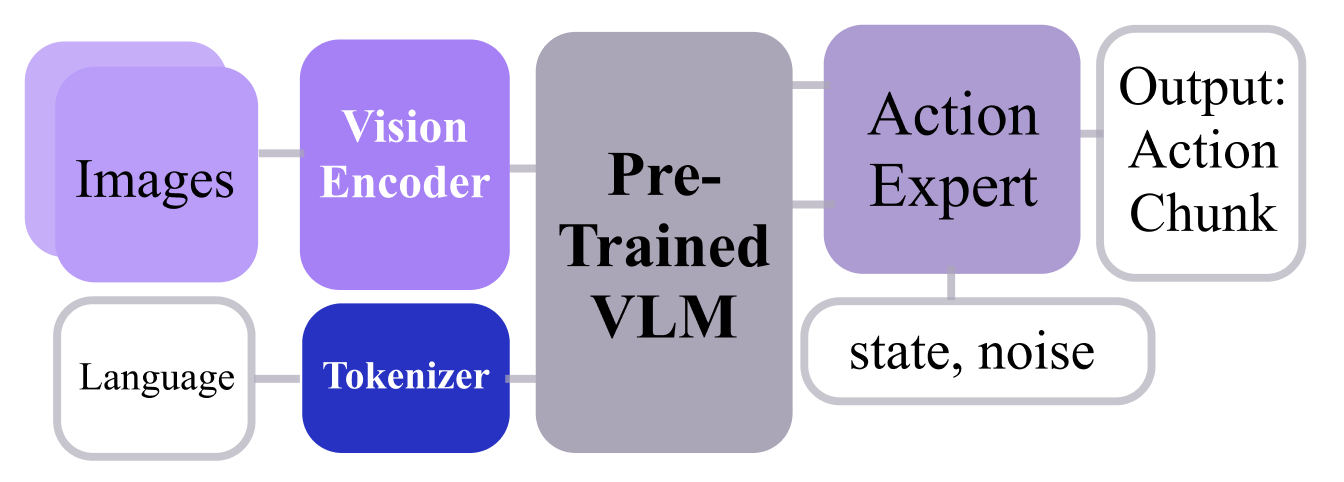}}
        \captionof{figure}{Overview of VLA models.}
        \label{fig:vla-overview}
    \end{minipage}
    \hfill
    \begin{minipage}{0.49\columnwidth}
    \centering
    \includegraphics[width=\columnwidth]{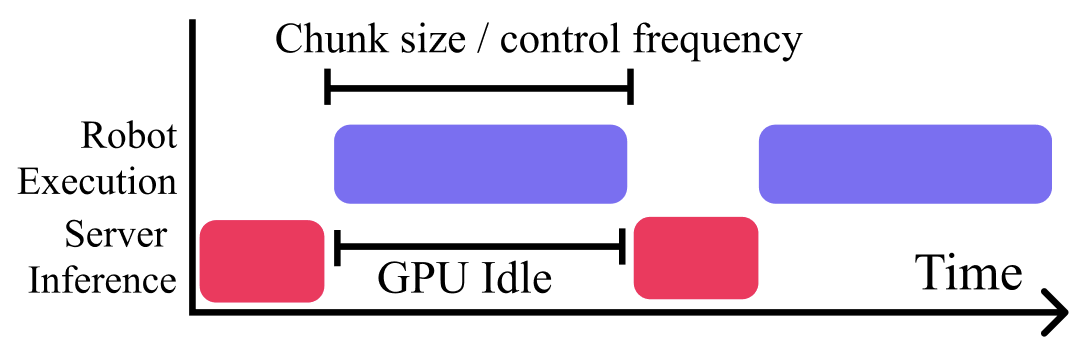}
        \captionof{figure}{Illustration of VLA inference and action execution.}
        \label{fig:sync-execution}
    \end{minipage}
\end{figure}

Vision-Language-Action (VLA) models \cite{black2025pi_, bjorck2025gr00t, wang2026vla} are an emerging paradigm in robotics that aim to learn generalizable robotics behavior. They leverage internet-scale vision encoders (ViT) and vision-language models (VLM) as backbones to understand image observations and user text-prompt instructions. A subsequent ``action expert'' (AE) stage generates a series of actions to achieve the text-prompted goal (see Figure \ref{fig:vla-overview}).

Unlike stand-alone VLMs, VLAs typically follow the ViT encoding and VLM prefill stages with a constant-step diffusion-based action policy \cite{black2025pi_, bjorck2025gr00t}, rather than a variable-length auto-regressive decode \cite{agrawal2023sarathi, yu2022orca}. 
The VLA ``action chunking'' paradigm entails that multiple action tokens are generated per inference pass, instead of a single decoded language token like with large-language models (LLMs) or VLMs. As a result, VLA workloads alternate between GPU inference and on-robot action execution (see Figure \ref{fig:sync-execution}).

While VLAs have improved robotic generalizability \cite{black2025pi_, intelligence2026pi}, we find they severely under-utilize GPU compute. Action chunking leaves the GPU idle between inference passes for $>$80\% of total task completion time while the robot executes generated actions. We also observe under-utilization during inference passes: streaming multiprocessor (SM) throughput averages 21-27\% across models, with the AE's utilization being the lowest of all three stages (5-19\%). Given the high cost of GPUs, VLA operators are economically motivated to maximize hardware utilization \cite{ye2024deep}. Since single-robot scenarios leave hardware idle, \textit{multi-client} VLA serving -- scheduling inference for multiple robotic clients on one device -- is a natural setting to study. This setting is most applicable to warehouse and factory deployments where robot fleets communicate with an on-premises server \cite{wen2025rosa}.

Multi-client serving enables techniques such as batching \cite{yu2022orca, agrawal2023sarathi} and stage co-execution \cite{xu2025nova, zhu2025nanoflow} to improve inference-time utilization, but these have not been characterized for VLAs. Prior LLM/VLM serving work \cite{yu2022orca, agrawal2023sarathi, zhu2025nanoflow, xu2025nova} does not account for the VLA's AE or action chunking, and prior multi-client VLA work \cite{wen2025rosa, pohland2026offloadoverloadplatformmeasurement} details neither the heterogeneous behavior of VLA stages nor stage co-execution.

We profile popular VLA models \cite{black2025pi_, bjorck2025gr00t, wang2026vla}, spanning key VLA architectural axes, to highlight fundamental challenges with these serving techniques applied to VLAs. Our analysis of batching shows that it alone cannot improve utilization across all stages due to the VLM's and ViT's compute-bound matrix-multiplication (GEMM) kernels. Furthermore, intuitions from LLM serving do not transfer directly to VLAs, since the latter cannot easily amortize the increased latency incurred from a batched VLM and ViT. Stage co-execution yields modest speedups as VLM GEMM kernels block AE kernel dispatching or slow substantially on partitioned SMs.

Therefore, in this characterization study, we make the following contributions:
\begin{itemize}[leftmargin=8pt]
    \item We explain VLA GPU compute under-utilization, both between and during inference passes, motivating the multi-client setting we characterize.
    \item We analyze VLA per-stage hardware utilization, power usage, and latency under batching, quantifying trade-offs and fundamental challenges for multi-client VLA serving.
    \item We explain the behavior of VLAs under stage co-execution, highlighting the limitations of co-execution techniques.
\end{itemize}

\section{Methodology}
\subsection{VLA Models}
We profile three popular open-source VLA models: \pimodel \cite{black2025pi_}, GR00T N1.7 \cite{bjorck2025gr00t}, and VLA-Adapter \cite{wang2026vla} using the compilation methods from their respective implementations. The models span key VLA architectural axes: parameter scale (0.5-3B), action-generation mechanism (flow-matching, L1-regression), and VLM-AE interface (full KV cache, single embedding, per-layer cross-attention), as summarized in Table \ref{tab:models}.

\begin{table}[t]
\centering
\small
\caption{Profiled VLA models. All are fine-tuned for the DROID
\cite{khazatsky2024droid} platform and run with two camera views.}
\label{tab:models}
\begin{tabular}{@{}llll@{}}
\toprule
 & \pimodel & GR00T N1.7 & VLA-Adapter \\
\midrule
VLM            & PaliGemma  & Cosmos-Reason2 & Prismatic \\
               & (Gemma)    & (Qwen3-VL)     & (Qwen2.5) \\
VLM/ViT params    & 3B & 2B          & 0.5B \\
AE  & flow-match. & flow-match. & L1-reg.  \\
AE params  & 300M       & 1B             & 100M \\
VLM-AE interf. & full KV    & single         & per-layer \\
               & cache      & embedding      & cross-attn. \\
\bottomrule
\end{tabular}
\end{table}

\subsection{Evaluation Methodology}
To reduce variability, our experiments report averages over 100 runs (fewer for NCU utilization measurements) after five warm-up runs. We input random images and use a short text prompt (40 tokens). Our measurements use CUDA synchronization. Utilization metrics come from NVIDIA Nsight Compute \cite{ncu}. 

\subsection{Hardware}
We profile each model on an NVIDIA RTX4090 and H100. The latter is a representative data-center GPU standard for LLM serving \cite{agrawal2023sarathi}. Although some VLAs have been adapted for deployment on embedded/onboard devices \cite{takagi2026anolevla}, we target larger GPUs for two reasons. First, server-side network and inference latencies are lower than onboard inference latency in most scenarios \cite{jiang2026fastirunvla}. Second, server-side GPUs enable larger models with further improved robot generalization \cite{hu2026bagelvla, intelligence2026pi}.

\section{Inter-Request Idleness}
\label{sec:inter-request-idleness}
\begin{figure}[h]
    \centering
    \begin{subfigure}{0.32\columnwidth}
        \centering
        \includegraphics[width=\linewidth]{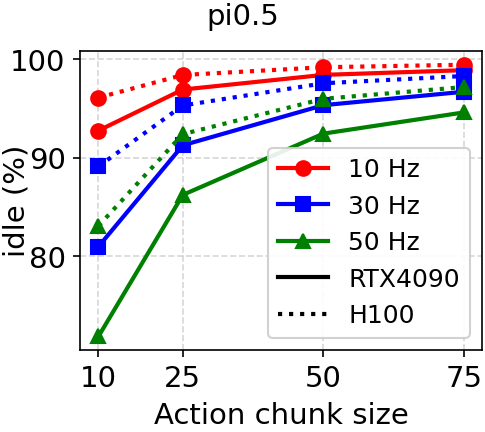}
    \end{subfigure}
    \begin{subfigure}{0.32\columnwidth}
        \centering
        \includegraphics[width=\linewidth]{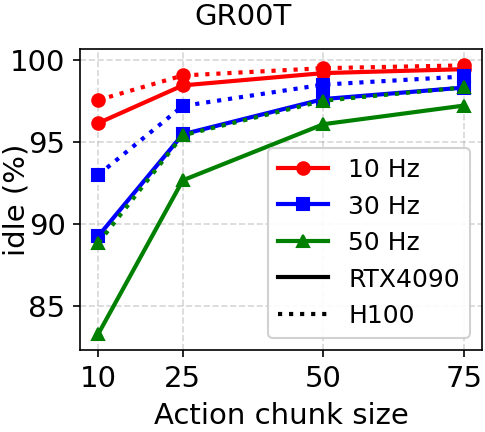}
    \end{subfigure}
    \begin{subfigure}{0.32\columnwidth}
        \centering
{}        \includegraphics[width=\linewidth]{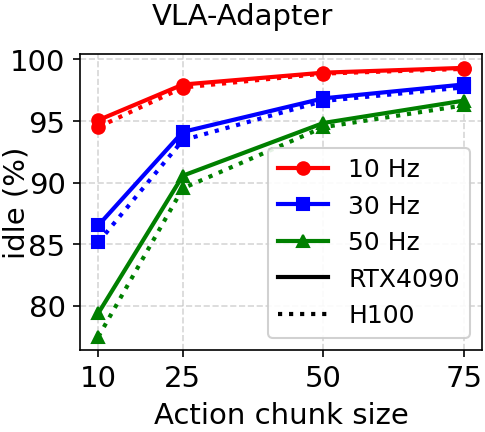}
    \end{subfigure}
    
    \caption{Percentage of time the GPU is idle, over different chunk sizes and control frequencies. Control frequency (in Hz) is the rate at which the robot executes actions.}
    \label{fig:idleness}
\end{figure}

Generating and executing action chunks creates alternating idle-busy periods for the GPU (see Figure \ref{fig:sync-execution}), which allows for multi-client scheduling \cite{pohland2026offloadoverloadplatformmeasurement}. In Figure \ref{fig:idleness}, we measure the effect that this action chunking has on GPU idleness. 

Figure \ref{fig:idleness} shows that the GPU is typically idle $>$80\% of the time. This idleness arises because generating an additional action token adds minimal inference latency: increasing the chunk size from 10 to 75 raises latency by $<$10\%. However, this same increase in chunk size corresponds to a 750\% increase in action execution time on the robot.

Therefore, action chunking leaves the GPU idle during the majority of each request cycle, even at high control frequencies and on consumer-grade GPUs. Because a single client occupies the GPU $\sim$15\% of the time, temporal interleaving alone can support $\sim$5-10 clients before the GPU's idle budget is consumed. Beyond that, additional clients' compute periods must overlap, so utilization gains require techniques that share the GPU under contention, such as batching and co-execution.

\section{Multi-request Batching}

\subsection{Utilization and Power Analysis}
\label{sec:batch_utilization}
\begin{figure*}[t]
    \centering
    \begin{subfigure}[c]{0.06\textwidth}
        \centering
        \includegraphics[width=\linewidth]{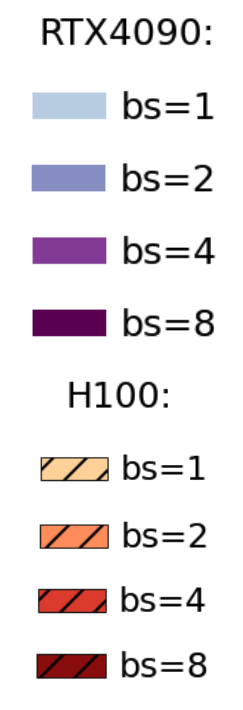}
    \end{subfigure}
    \begin{subfigure}[c]{0.3\textwidth}
        \centering
        \includegraphics[width=\linewidth]{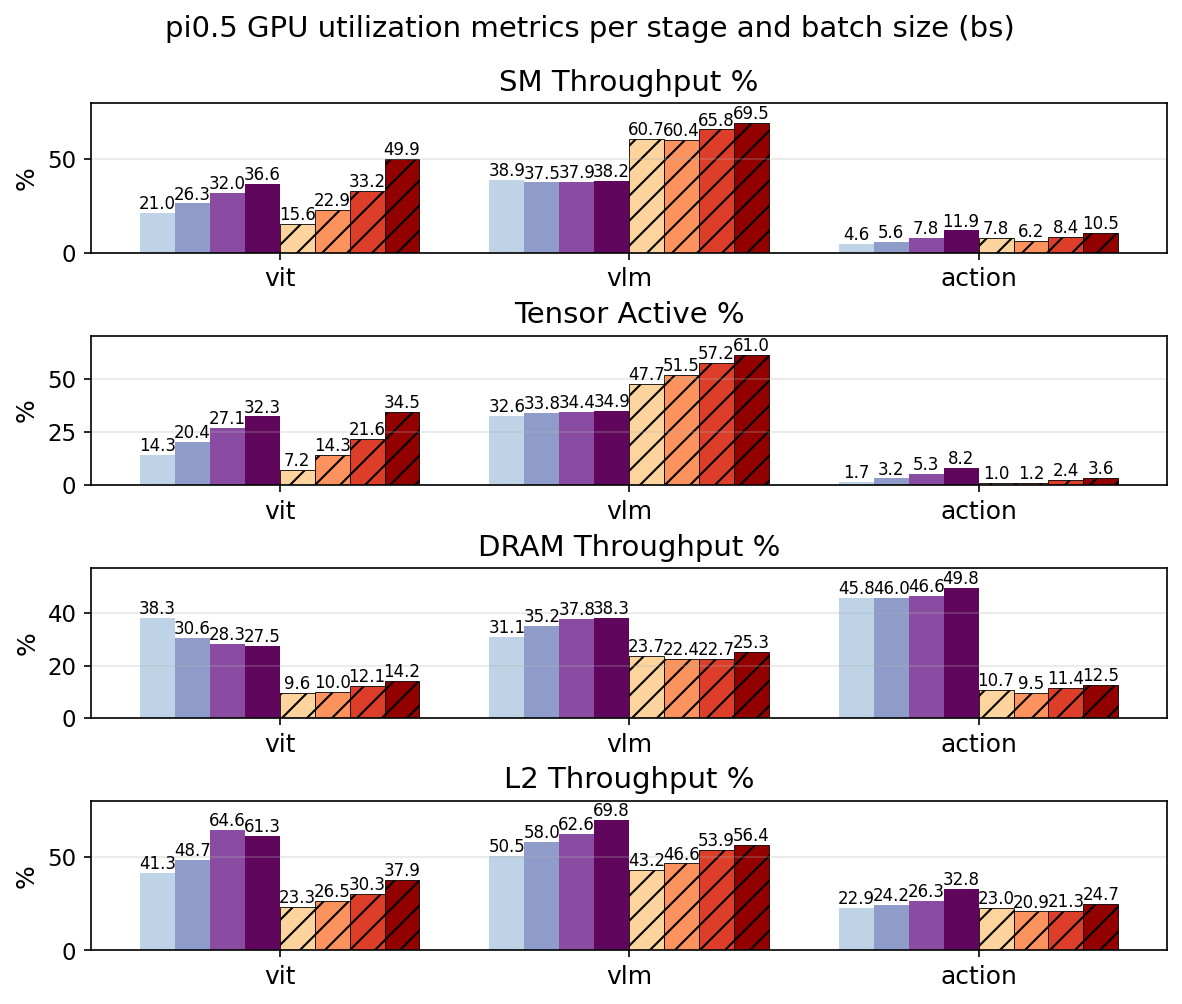}
    \end{subfigure}
    \begin{subfigure}[c]{0.3\textwidth}
        \centering
        \includegraphics[width=\linewidth]{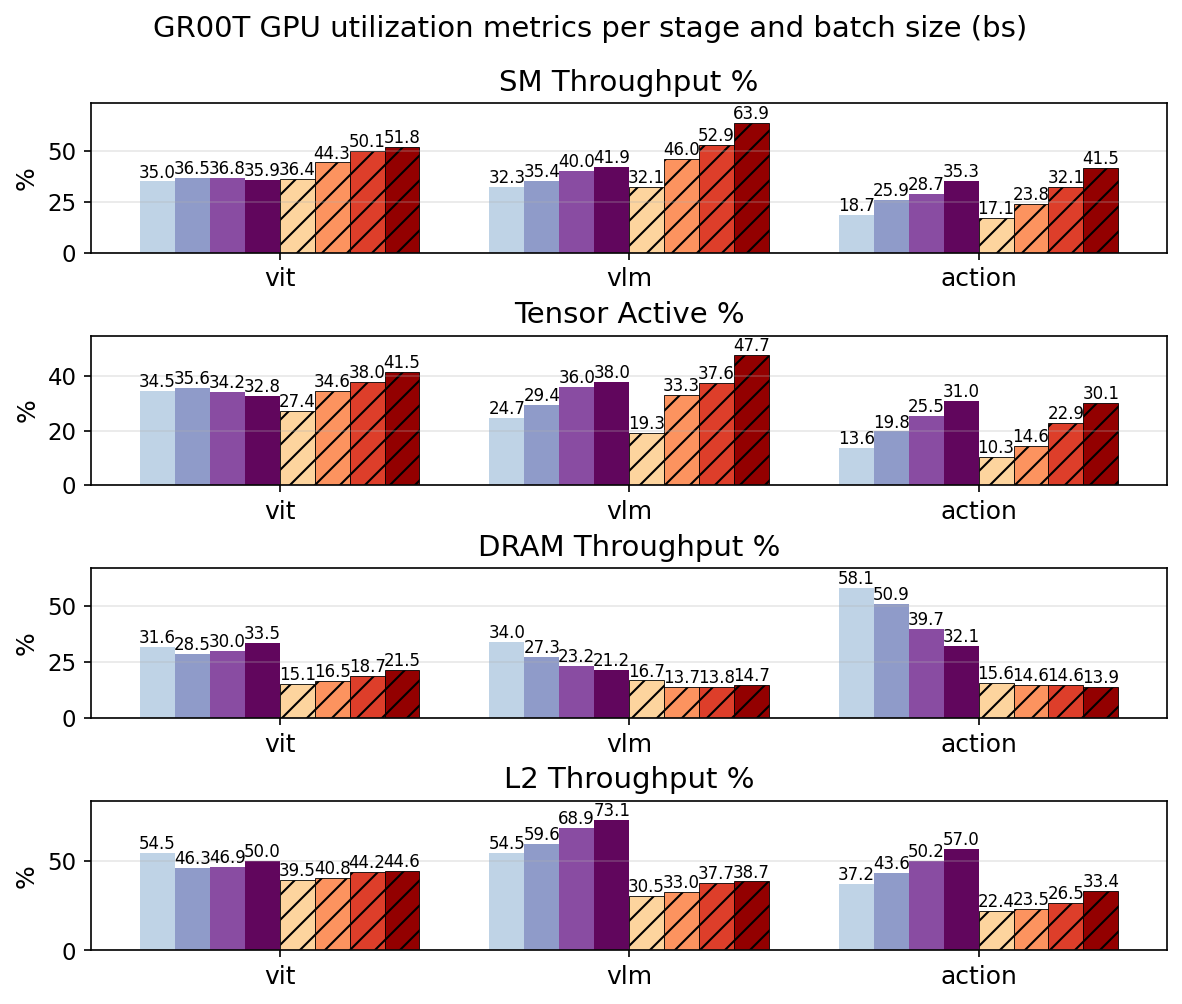}
    \end{subfigure}
    \begin{subfigure}[c]{0.3\textwidth}
        \centering
        \includegraphics[width=\linewidth]{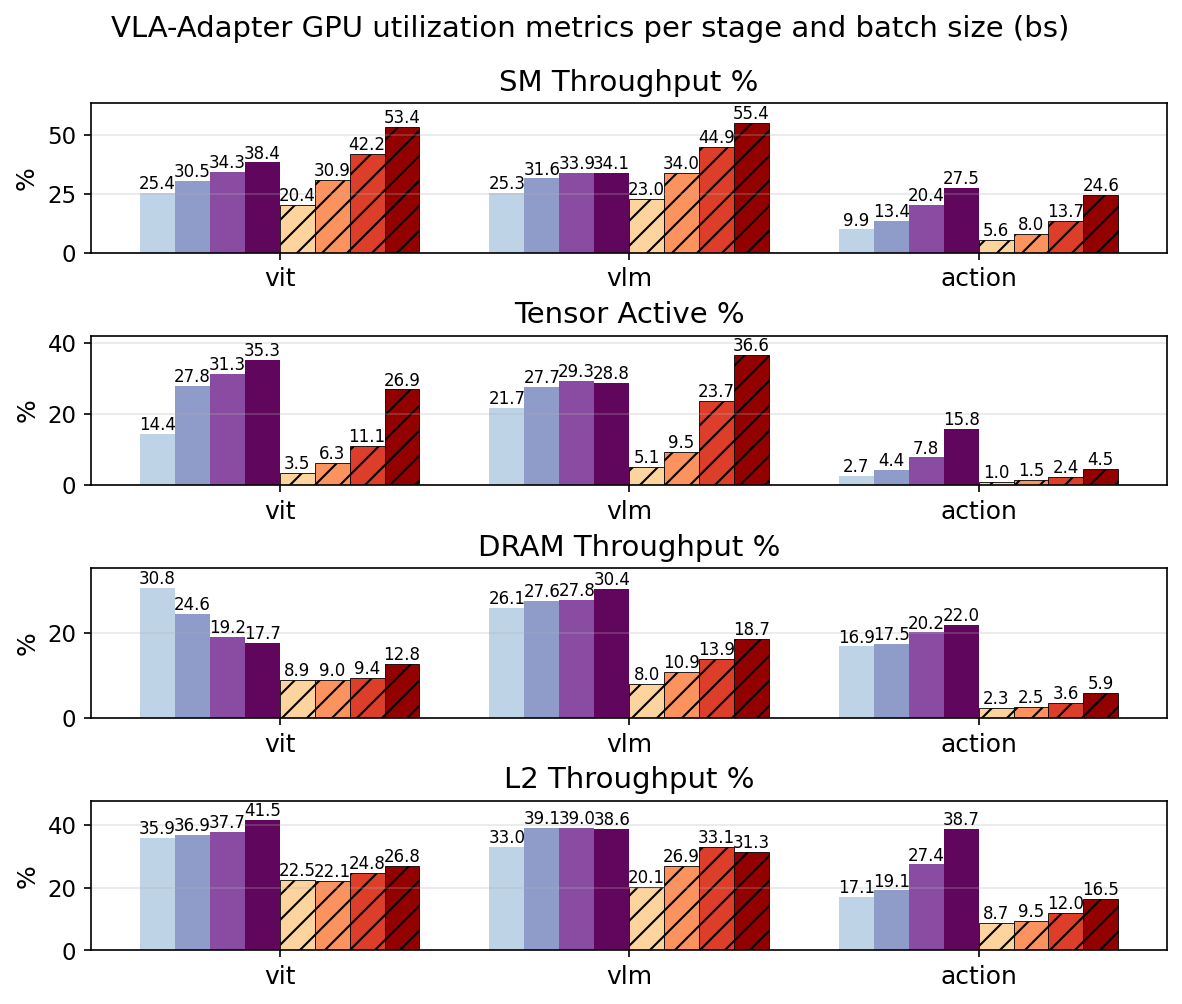}
    \end{subfigure}
    \caption{Utilization metrics over different batch sizes for $\pi_{0.5}$, GR00T and VLA-Adapter. Profiled on RTX4090 and H100.}
    \label{fig:batch_utilization}
\end{figure*}

\begin{figure}[h]
    \centering
    \begin{subfigure}{0.33\columnwidth}
        \centering
        \includegraphics[width=\linewidth]{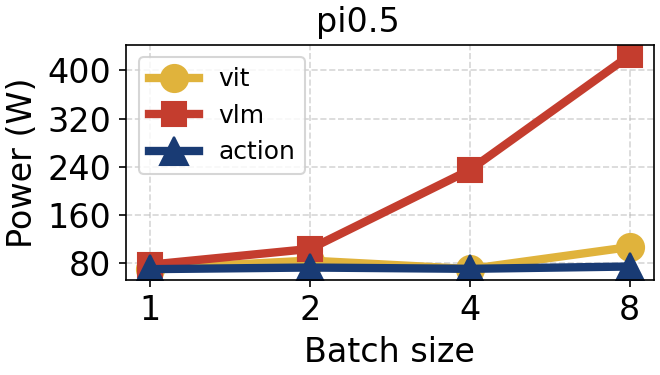}
    \end{subfigure}%
    \begin{subfigure}{0.33\columnwidth}
        \centering
        \includegraphics[width=\linewidth]{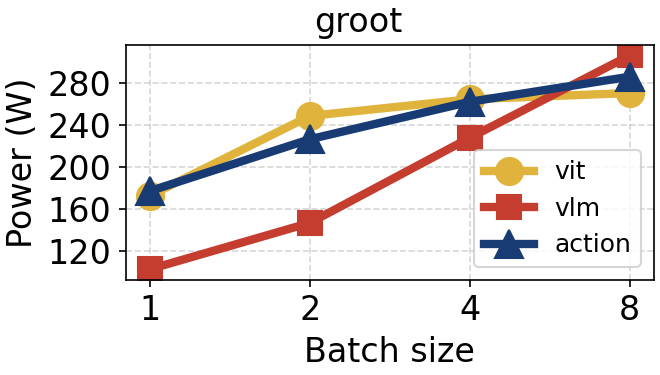}
    \end{subfigure}%
    \begin{subfigure}{0.33\columnwidth}
        \centering
        \includegraphics[width=\linewidth]{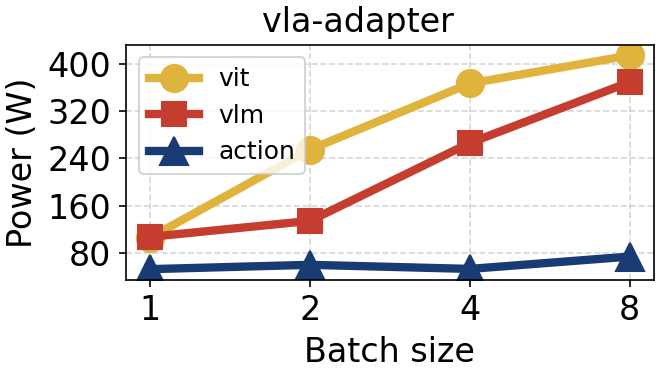 }
    \end{subfigure}
    
    \caption{Mean power per batch size of \pimodel, GR00T and VLA-Adapter on an RTX4090. To isolate the switching activity factor, we hold the GPU's clock frequency constant.}
    \label{fig:power_per_batch}
\end{figure}

GPUs are costly and consume substantial power. Consequently, VLA model operators have a strong economic incentive to maximize their utilization \cite{ye2024deep}. However, experiments reveal that VLA inference fails to saturate GPU compute, thereby raising the opportunity cost to serve these models. Batching, a fundamental technique in machine learning serving \cite{agrawal2024taming, yu2022orca}, combines input tensors so as to better exploit the GPU's parallel compute units and achieve higher utilization \cite{yu2022orca}. Figure \ref{fig:batch_utilization} and Figure \ref{fig:power_per_batch} measure the per-stage hardware utilization and power (respectively) of three VLA models as batch size increases. We make the following observations.

First, at a batch size of 1, the AE's compute utilization is low, with ``SM Throughput \%'' ranging from 4.6 to 18.7\%. Profiling \pimodel on an RTX4090 (representative case) shows that the AE's most time-consuming kernels are GEMM kernels, yet the average ``Tensor Active \%'' remains below 2\%. Two factors account for this. First, a weight-loading bottleneck: these kernels have high memory throughput (70-80\%) but low SM throughput (10\%). Second, the stage's key kernels are too small to saturate the GPU, as the AE computes attention over a short context, denoising a chunk of only 15 to 50 action tokens at once \cite{black2025pi_} while cross-attending to the VLM output. NCU profiling confirms this: the longest kernels run with low grid and block sizes. Because they issue too few instructions to overlap with outstanding memory requests, the weight-loading latency is exposed on the critical path.

Second, TCU utilization increases with batch size. All model stages exhibit some gain, but the effect is largest for the AE. The greatest TCU throughput gains originate from feed-forward network (FFN) kernels. These kernels grow linearly with batch size \cite{kaplan2020scaling}, allowing more warps to be scheduled and producing near-linear increases in TCU throughput.

Third, Figure \ref{fig:power_per_batch} shows that the power usage of the AE is characterized by the power required during the weight-loading bottleneck, which stays constant as batch size increases. For GR00T's AE, which has higher TCU utilization at a batch size of 1, batching the AE incurs a power overhead ($\sim$1.5$\times$ at a batch size of 4). 

Fourth, the VLM and ViT stages exhibit higher compute utilization than the AE. Attending to vision and text tokens yields a context length of 700 to 1000 tokens, which results in larger GEMM kernels with higher arithmetic intensity. VLM GEMMs are also larger than the AE's since their parameter count is typically one order of magnitude larger. Our profile traces show that the VLM's execution time is dominated by FFN GEMM kernels (65\% of time). Batching therefore has little effect on VLM compute utilization, as these time-dominating kernels already have high compute utilization at a batch size of one.

Fifth, L2 throughput increases with batch size for nearly all models and stages, since the larger GEMM kernels produced by batching exhibit greater data reuse through tiling \cite{nvidia_gemm_guide}.

Together, these results demonstrate that batching improves utilization primarily for the AE, while the backbone's utilization remains largely unchanged. Therefore, increasing overall utilization cannot be achieved through batching alone. 

\subsection{Latency Analysis}
\label{sec:batching_latency}
\begin{figure*}[h]
    \centering
    \begin{subfigure}{0.16\textwidth}
        \centering
        \includegraphics[width=\linewidth,]{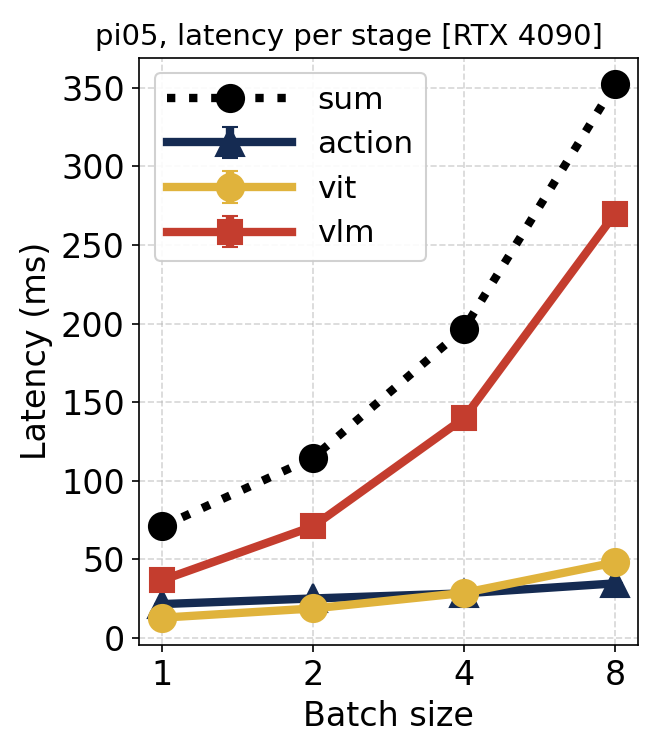}
    \end{subfigure}%
    \begin{subfigure}{0.16\textwidth}
        \centering
        \includegraphics[width=\linewidth]{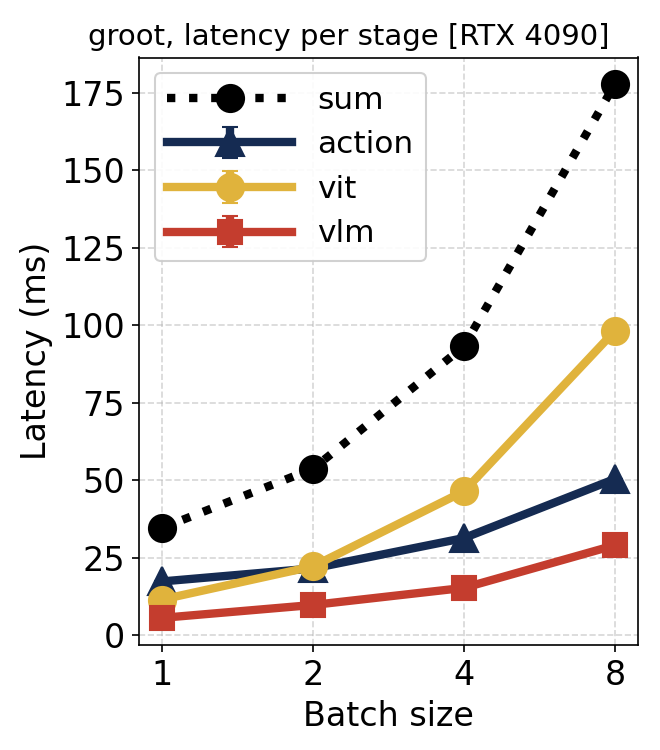}
    \end{subfigure}%
    \begin{subfigure}{0.16\textwidth}
        \centering
        \includegraphics[width=\linewidth]{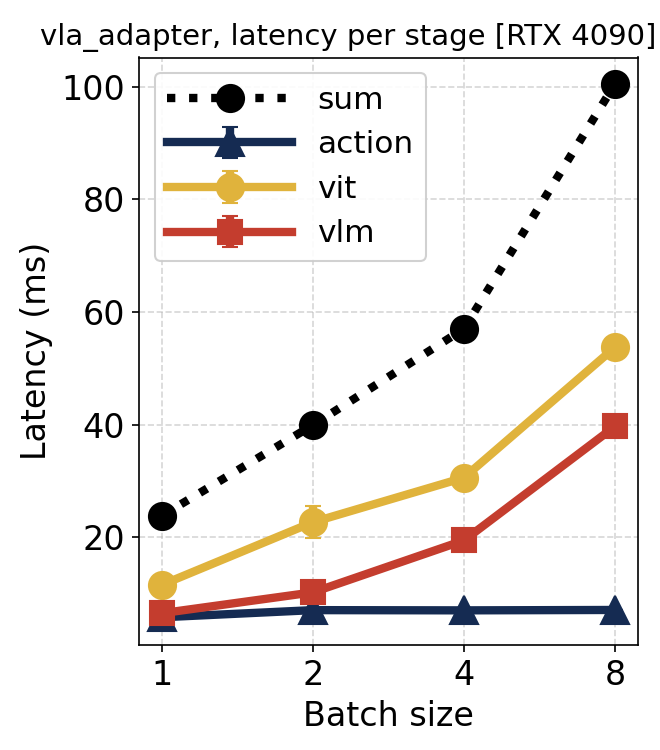}
    \end{subfigure}%
    \begin{subfigure}{0.16\textwidth}
        \centering
        \includegraphics[width=\linewidth]{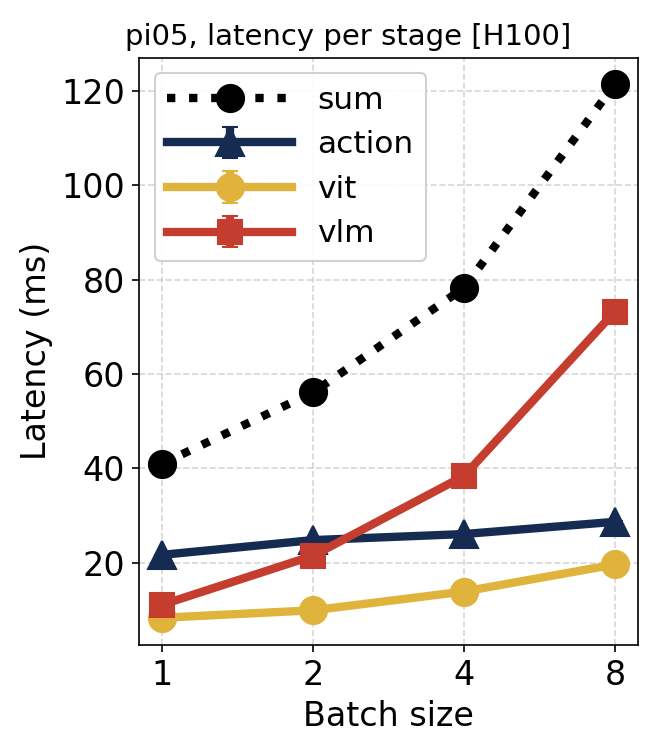}
    \end{subfigure}%
    \begin{subfigure}{0.16\textwidth}
        \centering
        \includegraphics[width=\linewidth]{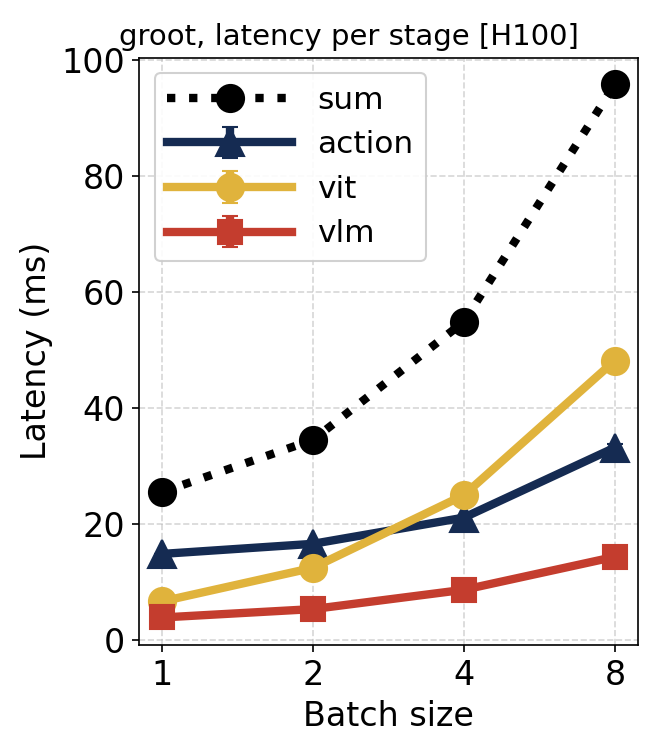}
    \end{subfigure}%
    \begin{subfigure}{0.16\textwidth}
        \centering
        \includegraphics[width=\linewidth]{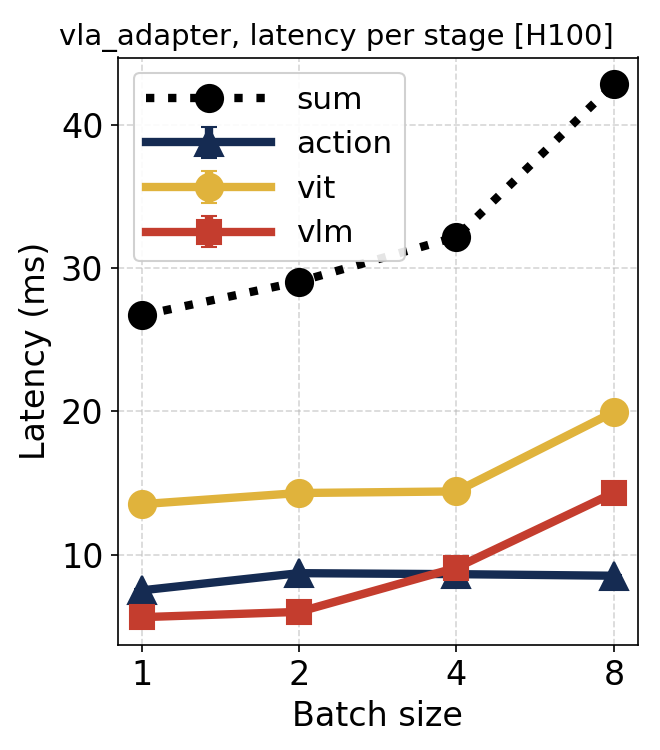}
    \end{subfigure}
    \caption{Per-batch latency for all models on RTX and H100 GPUs. Standard deviation bars are shown but are usually $<\pm$1ms.}
    \label{fig:latency_per_batch}
\end{figure*}

Batching can increase latency when the unbatched workload already saturates the GPU's compute units \cite{agrawal2024taming}. Because VLA response time is critical for safety and task completion \cite{jiang2026fastirunvla}, serving providers must balance utilization and throughput against latency. Yet, this trade-off has not been quantified for each VLA stage. In Figure \ref{fig:latency_per_batch} we measure the per-stage latency overhead of batching and make the following observations.

First, the AE batches with minimal latency overhead (1.15-1.33$\times$ at batch size 4, H100). The action-stage kernels perform relatively little computation and are memory-bound. Thus, once model weights are loaded, generating an additional batch of tokens adds little latency. The AE's low TCU utilization (Figure \ref{fig:batch_utilization}) means that large batches cause minimal extra stalling on the math pipeline.

Second, the VLM and ViT stages benefit less from batching, with much smaller throughput gains than the AE. In $\pi_{0.5}$'s VLM, $>$65\% of time is spent on dense FNN GEMM kernels (batch size 1), and $>$50\% of cycles stall on the math pipeline. Larger batches only deepen this compute bottleneck, raising latency.

Batching therefore affects VLA stages unevenly. The ViT and VLM process larger token counts, producing GEMM kernels that grow with batch size and dominate end-to-end latency (44-71\% at batch size 4). In LLM serving, comparable prefill costs are amortized over a long, variable-length decode stage \cite{agrawal2024taming}. However, VLAs cannot do this since action generation comes from a fixed-length AE. Furthermore, VLA input image tokens change on every pass, preventing prefill computation reuse via KV-cache optimizations. The VLM backbone is thus recomputed often and cannot be hidden by typical techniques used by LLM serving.

\section{Co-Execution}
GPU co-execution exploits the fact that a single application rarely saturates all of a GPU's heterogeneous resources (e.g., SMs, memory bandwidth, caches \cite{zhu2025nanoflow, xu2025nova}). Co-executing memory-bound and compute-bound applications has proven to be effective in other domains \cite{tan2022gpupool}, since their bottlenecks fall on different hardware, thus minimizing interference. Given the AE's under-utilized compute and the VLM stage's higher compute utilization (Section \ref{sec:batch_utilization}), we examine their co-execution as a candidate for stage-wise pipeline-parallel serving \cite{xu2025nova}.

Our experiments use standard co-execution methods: CUDA streams and CUDA green contexts (GC). The latter is an inter-SM partitioning method, which we tune for the highest throughput. Table \ref{tab:co-exec-latency} reports the throughput gains from these two methods.

\begin{table}[h!t]
  \caption{Throughput gains from co-executing VLM and AE relative to serial execution. For $\pi_{0.5}$, we use a PyTorch model with \texttt{torch.compile}. The green context split shows SMs partitioned for the VLM and AE respectively.}
  \centering
  \begin{tabular}{lllr}
    \hline
    Model- Hardware & CUDA Streams & GC (Split) \\
    \hline
    \pimodel - RTX4090 & 1.08$\times$ & 1.20$\times$ (88,40)\\
    \pimodel - H100 & 1.13$\times$ & 1.23$\times$ (88,44) \\
    GR00T - RTX4090 & 1.11$\times$ & 1.10$\times$ (32,96)\\
    GR00T - H100 & 1.02$\times$ & 1.04$\times$ (32,100)\\
    \hline
  \end{tabular}
  \label{tab:co-exec-latency}
\end{table}

From Table \ref{tab:co-exec-latency} we observe that the throughput of co-executed stages only marginally improves over serial execution (1.02-1.23$\times$). Below, we detail VLA stage co-execution challenges.

First, profile traces show that the VLM's large FFN and Query-Key-Value (QKV) GEMM kernels are a common source of action-stage stalls. These are the longest-running VLM kernels, with grid sizes large enough to launch a thread block per SM. Under multiple CUDA streams, a second kernel can only co-execute once the earlier kernel has dispatched all its thread blocks \cite{tan2022gpupool}, so AE kernel dispatching is blocked.

Second, while inter-SM partitioning improves throughput over CUDA streams, reduced compute units increase the latency of certain key kernels. The VLM's FFN and QKV projection GEMM kernels increase significantly in latency (50-100\% for $\pi_{0.5}$, 285-295\% for GR00T) given that these kernels are compute-intensive. Similarly, we observe that $\pi_{0.5}$'s AE sinusoidal time embedding kernels increase by over 225\% compared to their serial execution time, since these kernels use FP64 arithmetic. Additionally, the AE's GEMM kernels increase in latency (16-108\%) due to DRAM contention between the two stages (DRAM utilization increases by 33\%).

Third, action kernels are short-lived ($\sim$6µs), producing unpredictable overlap with long VLM kernels. Other co-execution methods therefore have relatively high overhead (40µs for preemption-based methods \cite{fan2025gpreempt}) or require re-implementing proprietary kernels. Finer-grained temporal SM repartitioning incurs high synchronization cost \cite{xu2025nova}.

Thus, despite their contrasting resource needs, standard methods fail to co-execute VLA stages efficiently. Under CUDA streams, the VLM's FFN and QKV kernels occupy most SMs and block AE kernel scheduling. Under SM partitioning, reduced SMs increase the latency of these compute-bound GEMM kernels in the VLM, and cause DRAM contention. Improving co-execution warrants more investigation.

\section{Conclusion}
We detail how fundamental serving techniques, batching and co-execution, affect key multi-client serving trade-offs for VLAs: latency, throughput, hardware utilization, and power. We find that batching alone cannot increase hardware utilization across all stages, as the VLM backbone is dominated by compute-bound GEMM kernels. We find that VLAs, unlike LLMs, cannot amortize the latency that batching the VLM and ViT stages introduces. Stage co-execution offers only modest gains since VLM GEMM kernels either block AE kernels or slow under SM partitioning. Short-lived AE kernels impede optimal co-execution. Our takeaways inform the trade-offs and challenges that multi-VLA serving systems must address.

\bibliographystyle{IEEEtran}
\bibliography{ref}

\end{document}